\documentclass[11pt,a4paper]{amsart}
\usepackage{amssymb}
\usepackage{xcolor}
\usepackage{graphicx}
\usepackage{verbatim}
\usepackage{enumerate}
\usepackage{hyperref}
\usepackage{float} 

\newcommand{\mr}{\mathrm}

\newcommand{\Mor}{\mathrm{Mor}}
\newcommand{\ra}{\rightarrow}
\newcommand{\M}{\mathcal{M}}
\newcommand{\ZZ}{\mathbb{Z}}
\newcommand{\mc}{\mathcal}
\newcommand{\bt}{\bullet}
\newcommand{\RR}{\mathbb{R}}
\newcommand{\F}{\mathbb{M}}
\newcommand{\g}{\mathfrak{g}}

\newcommand{\A}{\mathbb{A}}
\newcommand{\NN}{\widetilde{N}}
\newcommand{\dd}{\partial}
\newcommand{\PA}{I}
\newcommand{\til}{\widetilde}
\newcommand{\FF}{\mathbb{F}}
\newcommand{\MM}{\mathbb{M}}
\newcommand{\xra}{\xrightarrow}
\newcommand{\Lag}{\mathcal{L}}
\newcommand{\ul}{\mathsf}
\newcommand{\ui}{\mathsf{i}}
\newcommand{\up}{\mathsf{p}}
\newcommand{\uK}{\mathsf {K}}

\theoremstyle{remark}
\newtheorem{remark}{Remark}[section]
\theoremstyle{plain}

\theoremstyle{definition}

\newtheorem{example}{Example}
\newtheorem{theorem}{Theorem}
\newtheorem{assumption}[remark]{Assumption}

\begin{document}

\title[]{Two field-theoretic viewpoints on the Fukaya-Morse $A_\infty$ category}
\author{Olga Chekeres}

\address{University of Connecticut, Department of Mathematics, Storrs, CT 06269, USA}
\email{olga.chekeres@uconn.edu}

\author{Andrey Losev}

\address{Wu Wen-Tsun Key Lab of Mathematics, Chinese Academy of Sciences, 
USTC, No.96, JinZhai Road Baohe District, Hefei, Anhui, 230026, P.R.China}
\address{National Research University Higher School of Economics \\
Laboratory of Mirror Symmetry, NRU HSE, 6 Usacheva str., Moscow,  Russia, 119048
}
\address{Federal Science Centre ``Science Research Institute of System Analysis at Russian Science Academy'' (GNU FNC NIISI RAN),  Nakhimovskiy pr. 36-1, Moscow, Russia, 117218}
\email{
aslosev2@yandex.ru
}

\author{Pavel Mnev}

\address{University of Notre Dame, Notre Dame, IN 46556, USA}
\address{St. Petersburg Department of V. A. Steklov Institute of Mathematics of the Russian Academy of Sciences, 
27 Fontanka, St. Petersburg, Russia, 191023}
\email{pmnev@nd.edu}

\author{Donald R. Youmans}

\address{Albert Einstein Center for Fundamental Physics,
Institute for Theoretical Physics,
University of Bern, Hochschulstrasse 6, 3012 Bern, Switzerland}
\email{youmans@itp.unibe.ch}

\thanks{The work of A. S. Losev is partially supported by Laboratory of Mirror Symmetry NRU HSE, RF Government grant, ag. 
N\textsuperscript{\underline{o}}  14.641.31.0001.
The work of D. R. Youmans is supported by the NCCR SwissMAP of the Swiss National Science Foundation.}

\begin{abstract}
We study an enhanced version of the  Morse degeneration of Fukaya $A_\infty$ category with higher compositions given by numbers of gradient flow trees. The enhancement consists in allowing morphisms from an object to itself to be chains on the manifold.
Higher compositions correspond to counting Morse trees passing through a given set of chains. We provide two viewpoints on the construction and on the proof of the $A_\infty$ relations for the composition maps. One viewpoint is via an effective action for the $BF$ theory  computed in a special gauge. The other is via higher topological quantum mechanics.
\end{abstract}


\maketitle

\tableofcontents

\section{Introduction}
In 
\cite{Fukaya}
Fukaya introduced an $A_\infty$  
category whose objects are Lagrangian submanifolds in a 
symplectic
manifold, morphisms are  formal linear combinations of intersection points
of these manifolds and higher compositions are 
 defined in terms of the count of
 holomorphic\footnote{By an abuse of language, here and below we say ``holomorphic'' while meaning ``pseudoholomorphic,'' with respect to a compatible almost complex structure on the symplectic manifold.} disks bounded by these manifolds.\footnote{
 We refer the reader to \cite{FOOO} and \cite{Seidel} for details and to \cite{Auroux} for an introduction to the subject.}
If the 
symplectic
manifold is a cotangent bundle $T^*X$ and the Lagrangian submanifolds can be projected onto the base $X$, then $N$ different 
Lagrangian submanifolds $L_a$ (assuming they are exact)  may be described
by a set of $N$ generating functions $F_a$, so that $L_a$'s 
are given by equations 
$$
p_i=\frac{\partial F_a}{\partial X^i},
$$
where $p_i$ are coordinates in the fiber dual to  the coordinates $X^i$ on the base.
Then intersection points 
of Lagrangian submanifolds project to the 
 critical points $P_{\alpha, ab}$  
 of the difference of generating functions
$$
d(F_a-F_b)=0.
$$
If we consider the rescaling $F_a \rightarrow \epsilon F_a$, then the projection to the base of a holomorphic disk tends to a ribbon tree formed by
gradient trajectories of differences of generating functions. Counting such ribbon trees 
determines the composition maps of
an $A_\infty$ 
category. 
We call this structure 
``Fukaya-Morse $A_\infty$ category,'' 
to distinguish it from the ``usual'' Fukaya category of Lagrangians in a symplectic manifold given by 
counts of holomorphic disks.
We refer to the original papers \cite{Fukaya,FO} and also to \cite{KS} for details.

What is missing in this construction is the description of
 morphisms from an object to itself.
We would like to fill in this gap and present an $A_\infty$ 
structure containing such morphisms. We will present a construction and give 
two different proofs
of the $A_\infty$ relations. 
 Neither of these proofs will involve holomorphic disks.
 
\subsection{Summary of results}
The novelty of this paper with respect to existing literature is the following three points.
\begin{enumerate}[(A)]
\item Section \ref{sec 2}: Enhancement of the Fukaya-Morse $A_\infty$ category by a model for endomorphisms of an object: 
$\Mor(F_a,F_a)=C_\bt^\mr{sing}(X,\ZZ)$ -- singular chains on $X$. Higher compositions are defined by 
an
enumerative geometric problem: counting
trees made of gradient trajectories in $X$, connecting a given set of critical points and passing through a given set of chains.
\item Section \ref{sec 3}:  A description of the data of the enhanced Fukaya-Morse $A_\infty$ category in terms of an effective action for a topological gauge theory (the $BF$ theory), computed in a particular gauge. In this framework, structure relations of $A_\infty$ category follow from the Batalin-Vilkovisky master equation for the effective action.
\item Section \ref{sec 4}: A description of the data of the enhanced Fukaya-Morse $A_\infty$ category in terms of higher topological quantum mechanics -- a 
differential form $I$ on the moduli space of metric trees. 
In this language, $A_\infty$ composition maps are periods of the form $I$ and the $A_\infty$ relations follow from Stokes' theorem on the moduli space of trees and from the factorization property of $I$ at the boundary of the moduli space. This proof is a 1D analog of the proof of the 
WDVV 
equation in cohomological field theory, see \cite{KM}.
\end{enumerate}

\section{Enhanced Fukaya-Morse $A_\infty$ category, construction and examples} \label{sec 2}
\subsection{Definition of the ``enhanced'' Fukaya-Morse $A_\infty$ category}
\label{ss: definition of F}

In this section we describe an enhanced version
of  the
Fukaya-Morse 
$A_\infty$ pre-category\footnote{ ``$A_\infty$ pre-category,'' in the terminology of \cite[section 4.3]{KS}, refers to higher composition being defined only for sequences of morphisms connecting ``transversal'' sequences of objects (the class of transversal sequences of objects is a part of data of an $A_\infty$ pre-category). The notion of a ``topological $A_\infty$ category'' from \cite{FO} contains a similar caveat: compositions are defined for a Baire subset of sequences of objects. We do not use either of those terms for our enhancement (rather we say ``partially defined'') since our transversality condition is on a sequence of morphisms, not just on the objects connected by those morphisms.
} 
of \cite{Fukaya,FO,KS}. More precisely, we consider a full subcategory of what is called in \cite[section 6.2]{KS} ``Morse $A_\infty$ pre-category of smooth functions.''  The enhancement is  given by a model for morphisms from an object to itself. The resulting structure is a \emph{partially defined} $A_\infty$ category: higher composition maps are defined for a dense subset of sequences of input morphisms. 

\indent
\textbf{Objects.} We consider the $A_\infty$ category $\mathbb{F}$ (with partially defined compositions) 
where the set of objects is a collection of smooth functions $\{F_1,\ldots,F_N\}$  in general position on a closed Riemannian manifold $X$.  In particular, we assume that differences $F_a-F_b$ are Morse-Smale functions\footnote{Recall that a function $F$ on $X$ is Morse if it has isolated critical points and its Hessian at each critical point is nondegenerate. $F$ is said to be Morse-Smale if additionally for any pair of critical points $P,Q$ the stable (ascending) manifold of $P$ and unstable (descending) manifold of $Q$ intersect transversally.} for $a\neq b$.

\textbf{Morphisms.} For $a\neq b$, the 
 morphisms from object $F_a$ to object $F_b$ are Morse chains of $F_a-F_b$, 
\begin{equation}
 \mr{Mor}(F_a,F_b)\colon =MC^\bullet(F_a-F_b)=\mr{span}_\mathbb{Z}(\mr{Crit}(F_a-F_b))
\end{equation}
-- formal linear combinations of critical points of $F_a-F_b$, taken with cohomological grading by Morse coindex of a critical point. We take the space of morphisms between $F_a$ and itself to be 
 the complex of smooth singular chains on $X$, 
\begin{equation}
 \mr{Mor}(F_a,F_a)= C^\mr{sing}_\bullet(X,\mathbb{Z}),
\end{equation}
where we consider chains with cohomological grading by codimension. We remark that there are other possible models for endomorphisms, see Section \ref{s: models for End}.
 
\textbf{Composition maps.} 
We have  higher composition maps
{\small 
\begin{multline}\label{m source, target}
m\colon \Mor(F_{a_1},F_{a_1})^{\otimes k_1}\otimes \Mor(F_{a_1},F_{a_2})\otimes \Mor(F_{a_2},F_{a_2})^{\otimes k_2}\otimes \Mor(F_{a_2},F_{a_3})\otimes\cdots\\ 
\otimes \Mor(F_{a_{r-1}},F_{a_r})\otimes \Mor(F_{a_r},F_{a_r})^{\otimes k_r}\quad \rightarrow \quad \Mor(F_{a_1},F_{a_r})
\end{multline}
}
defined as follows.
\begin{enumerate}[(i)]
\item For $r\geq 2$:
{\small 
\begin{multline}\label{mu}
m(\{Z_{\alpha,a_1}\}_{\alpha=1}^{k_1};[P_{a_1a_2}];\{Z_{\alpha,a_2}\}_{\alpha=1}^{k_2};[P_{a_2a_3}];\cdots;[P_{a_{r-1}a_r}];\{Z_{\alpha,a_r}\}_{\alpha=1}^{k_r})=\\
\!\!\!\!\!\!\!
\sum_{P_{a_1 a_r}\in \mr{Crit}(F_{a_1}-F_{a_r})} \# \mathcal{M}(\{Z_{\alpha,a_1}\}_{\alpha=1}^{k_1};P_{12};\{Z_{\alpha,a_2}\}_{\alpha=1}^{k_2};P_{a_2a_3};\cdots;P_{a_{r-1}a_r};\{Z_{\alpha,a_r}\}_{\alpha=1}^{k_r};P_{a_1 a_r})\cdot [P_{a_1 a_r}].
\end{multline}
}
\item For $r =1$, we define $m\colon  \Mor(F_a,F_a)^{\otimes k}\ra \Mor(F_a,F_a)$ to be
\begin{equation}\label{m with r=1}
m(Z_1)=\dd Z_1,\quad m(Z_1,Z_2)=Z_1\cap Z_2, \quad m(Z_1,\ldots,Z_k)=0 \;\mr{for}\; k\geq 3.
\end{equation}
\end{enumerate}
Here: 
\begin{itemize}
\item $P_{ab}$ are critical points of $F_a-F_b$; $[P_{ab}]$ denotes the the critical point seen as a vector in $\Mor(F_a,F_b)$.
\item $Z_{\alpha,a}$ are chains on $X$.  
\item $\mathcal{M}(\cdots)$ is the moduli space of ``Morse trees'' -- binary rooted embedded trees in $X$ with 1-valent vertices (``leaves'') at $P_{a_i a_{i+1}}$ and root at $P_{a_1a_r}$. Each edge carries a bi-index $(a,b)$ with $a\neq b$ and is a gradient trajectory of $F_a-F_b$. A 1-valent vertex $P_{a_i a_{i+1}}$ is adjacent to a $(a_i, a_{i+1})$-edge, the root is adjacent to a $(a_1, a_r)$-edge. At a 3-valent vertex, the incoming edges are $(a,b)$, $(b,c)$ and the outgoing is $(a,c)$, for some $a,b,c$. We are only considering Morse trees subject to the extra condition that each chain $Z_{\alpha,a}$ intersects an $(a,b)$- or $(b,a)$-edge of the tree, for some $b$.  Moreover, the order of chains $Z_{\alpha,a_i}$ (as inputs of $m$) for a given $i$ should be the order in which intersections with chains $Z_{\alpha,a_i}$ occur, as one goes along the path in the tree connecting $P_{a_{i-1} a_i}$ to $P_{a_i a_{i+1}}$ (in particular, $Z_{\alpha,a_1}$ are along the path from the root to $P_{a_1a_2}$ and $Z_{\alpha,a_r}$ are along the path from $P_{a_{r-1}a_r}$ to the root).
\item $\#\mathcal{M}(\cdots)$ is the number of Morse trees if the moduli space is zero-dimensional, and zero otherwise.
\item  The composition (\ref{mu}) is defined under certain transversality assumption -- Assumption \ref{assump: transversality} below -- on the input morphisms, i.e., on critical points $P_{ab}$ and chains $Z_{\alpha,a}$. 
Also, in (\ref{m with r=1}) the binary operation is the  intersection of chains and we are assuming that chains $Z_1,Z_2$ are intersecting transversally. These assumptions constitute the reason why compositions in $\FF$ are partially defined.  
\item Subtlety: in (\ref{mu}) we are implicitly assuming $a_1\neq a_r$. Compositions in the case $a_1=a_r$ can be obtained from cyclicity, see below.
\end{itemize}

The composition map (\ref{m source, target}) with $n$ inputs has cohomological degree $2-n$.


\textit{Differentials on $\Mor$.} Composition maps with one input morphism are the differentials on morphism spaces. In particular, $m\colon \Mor(F_a,F_b)\ra \Mor(F_a,F_b)$ is the Morse differential for $a\neq b$ and 
the boundary operator on chains for $a=b$, by the definition above. 


\textit{Duals/cyclicity.} One has a natural pairing on morphism spaces 
$$\langle \,,\,\rangle\colon \Mor(F_a,F_b)\otimes \Mor(F_b,F_a)\ra \ZZ.$$
For $a=b$, this is the (signed) count of points in the intersection of two chains.\footnote{We are assuming the intersection to be transversal, thus the pairing is also partially defined.} For $a\neq b$, we set $\langle [P],[P'] \rangle=\delta_{P\, P'}$ -- using the fact that critical points $P$ of $F_a-F_b$ are simultaneously critical points of $F_b-F_a$. Thus, one can say that the category $\mathbb{F}$ has duals for morphisms, $\Mor(F_a,F_b)^\vee=\Mor(F_b,F_a)$. Using this, one can define \emph{cyclic $A_\infty$ compositions}
\begin{equation}\label{cyclic composition}
c(x_0,\ldots,x_p)\colon=\langle x_0, m(x_1,\ldots,x_p)\rangle
\quad \in \ZZ,
\end{equation}
where $\{x_i\}$ are morphisms in $\mathbb{F}$ such that $\mr{target}(x_i)=\mr{source}(x_{i+1})$ where index $i$ is 
in $\ZZ_{p+1}$.
Cyclic compositions coincide with the numbers $\#\M$ appearing in (\ref{mu}) and are invariant under cyclic permutation of inputs.

\textit{Case $a_1=a_r$.} In (\ref{mu}) we were implicitly assuming $a_1\neq a_r$. In the case $a_1=a_r$, one can recover the composition map $m$ from cyclicity -- from invariance of (\ref{cyclic composition}) under cyclic permutations of inputs. For example, the map $m\colon \Mor(F_1,F_2)\otimes \Mor(F_2,F_1)\ra \Mor(F_1,F_1)$ sends $[P]\otimes [Q]$ (with $P,Q$ two critical points of $F=F_1-F_2$) to 
$\mr{Unstab}_P\cap \mr{Stab}_Q$
-- 
the intersection of the unstable manifold of $P$ and stable manifold of $Q$ (with respect to the Morse function $F$).

 Restricting (\ref{mu}) to the case when there are no input morphisms in $\Mor(F_a,F_a)$ (no chains $Z_{\alpha,a}$), one gets the usual Fukaya-Oh construction of \cite{Fukaya,FO}.

\begin{theorem}\label{thm 1}
The composition maps $m$ satisfy the $A_\infty$ relations:
if $x_1,\ldots,x_n$ is a collection of morphisms  in $\mathbb{F}$ satisfying $\mr{target}(x_i)=\mr{source}(x_{i+1})$ for $i=1,\ldots,n-1$, 
then we have
\begin{equation}\label{A_infty rel}
\sum_{r\geq 0,\;s\geq 1,\; r+s\leq n} \pm m(x_{1},\ldots,x_{r},m(x_{r+1},\ldots,x_{r+s}),x_{r+s+1},\ldots,x_{n})=0.
\end{equation}
\end{theorem}

\subsection{Examples and pictures}
While the definition above looks a bit heavy, the concept behind it is intuitively clear and can be illustrated by 
 simple examples.

\begin{example} 
The composition map 
$$m\colon \Mor(F_1,F_2)\otimes \Mor(F_2,F_3)\ra \Mor(F_1,F_3)$$ (no singular chains involved) is given by counting trees made of gradient trajectories like this:
\begin{figure}[H]
$$
\vcenter{\hbox{ \includegraphics[scale=0.8]{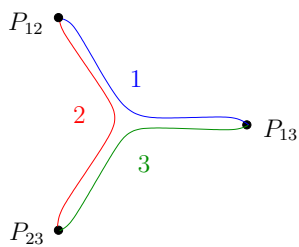} }} 
$$
\caption{Morse tree}
\label{fig1}
\end{figure}

Here 
is a more detailed picture explaining the origin of the tree from the Fukaya category of Lagrangians (of the form $L_a=\mr{graph}(\epsilon\, dF_a)$) in $T^*X$ with composition maps given by counting  holomorphic disks:
\begin{figure}[H]
$$\vcenter{\hbox{ \includegraphics[scale=0.7]{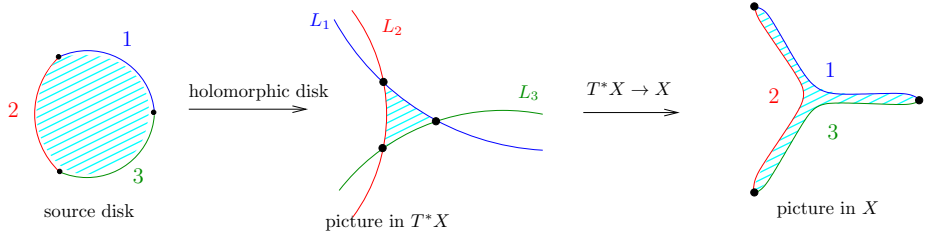} }} $$
\caption{Morse tree in $X$ as a  projection of a holomorphic disk in $T^*X$ (in $\epsilon\ra 0$ limit).}
\label{fig2}
\end{figure}

And here 
is a modification of 
this picture
for the composition map 
$$m\colon \Mor(F_1,F_2)\otimes \Mor(F_2,F_2)\otimes \Mor(F_2,F_3)\ra \Mor(F_1,F_3)$$ 
involving a singular chain $Z\in \Mor(F_2,F_2)$ on $X$ (respectively, on a Lagrangian in $T^*X$):
\begin{figure}[H]
$$\vcenter{\hbox{ \includegraphics[scale=0.7]{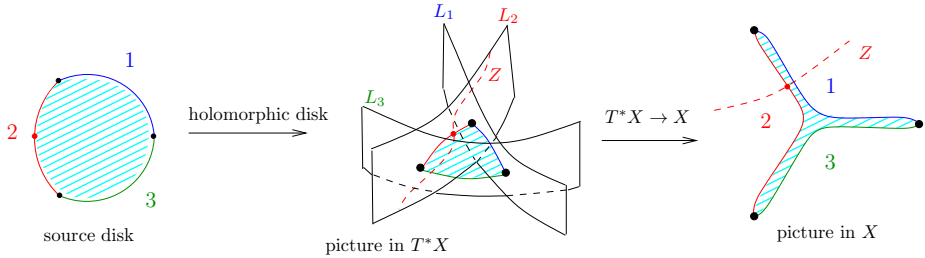} }} $$
\caption{Holomorphic disk and Morse tree intersecting a chain.}
\label{fig3}
\end{figure}
\end{example}

\begin{example}  
The composition map
$$m\colon \Mor(F_1,F_1)^{\otimes k}\otimes \Mor(F_1,F_2)\otimes \Mor(F_2,F_2)^{\otimes l}\ra \Mor(F_1,F_2)$$ 
is given by counting gradient trajectories of $F_1-F_2$ intersecting a given set of chains (dashed curves in the picture):
\begin{figure}[H]
$$ \vcenter{\hbox{ \includegraphics[scale=0.8]{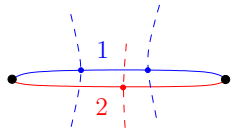} }}$$
\caption{Gradient trajectories passing through a set of chains.}
\label{fig4}
\end{figure}
The dimension of the corresponding moduli space of gradient trajectories is 
$$\dim\mathcal{M}=\mr{ind}(P)-\mr{ind}(P')-1-\sum_{\alpha=1}^k (\mr{codim}(Z_{\alpha,1})-1)-\sum_{\beta=1}^l (\mr{codim}(Z_{\beta,2})-1).$$ The balancing condition on indices of critical points and dimensions of chains, necessary for the morphism to be nonzero, is $\dim\mathcal{M}=0$.
\end{example}

\begin{example}[Example of an $A_\infty$ relation] 
\label{ex: heart-shaped sphere}
Let $X$ be the 2-sphere, $F_{1,2}$ two functions such that $F_1-F_2$ is the height function of the heart-shaped sphere. Let $P,Q$ be the two critical points of $F_1-F_2$ of index 2, $R$ the critical point of index 1 and $S$ the index 0 critical point, and $Z$ an embedded interval, as in the picture below.    

\begin{figure}[H]
$$ \vcenter{\hbox{ \includegraphics[scale=1]{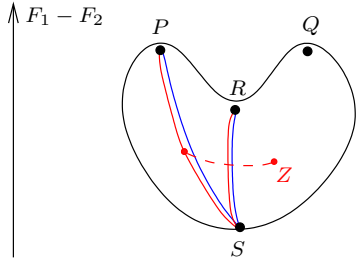} }}$$
\caption{Two first terms in (\ref{A_infty rel on heart}): gradient trajectories passing through a chain/its boundary.}
\label{fig5}
\end{figure}
One has the relation
\begin{equation}\label{A_infty rel on heart}
m(\underbrace{d_\mr{Morse}[P]}_{[R]},Z)+m([P],\partial Z)+d_\mr{Morse}m([P],Z)=0.
\end{equation}
The first two terms give $S$ with opposite signs and the third term vanishes (since the moduli space $\mathcal{M}(P,Z,\widetilde{P})$ is 1-dimensional if $\widetilde{P}=S$
and empty if $\widetilde{P}$ is any other critical point).  This is the $A_\infty$ relation (\ref{A_infty rel}) 
with input morphisms $x_1=[P]$, $x_2=Z$.
\end{example}

\subsection{Morse contraction}\label{sec: Morse contraction} For the next example we need to recall 
the construction of 
\emph{Morse contraction} from the de Rham complex to the Morse complex from \cite[section 6.5]{KS}; it will also be a key ingredient for the following sections.
 
For a given Morse-Smale function $F$ on $X$, let $v$ be the corresponding gradient vector field. Consider the expression
\begin{equation}\label{U}
U(t,dt)= e^{-t \mc{L}_v} (1-dt\; \iota_v).
\end{equation}
This is a nonhomogeneous differential form on $(0,+\infty)$ with values in linear operators acting on differential forms on $X$.\footnote{ The operator
(\ref{U}) is the evolution operator of Morse topological quantum mechanics, understood as a differential form on the moduli space of metric intervals, 
see \cite{Witten_Morse}, \cite{FLN} and Section \ref{ss: HTQM idea} below.
} 
Here and below by $e^{-t \mathcal{L}_v}$ we mean the pullback by the flow of $v$ in time $-t$.


\textbf{Non-regularized Morse contraction.}

Consider the following triple of maps.
\begin{itemize}
\item The inclusion $i\colon MC^j(F,\RR)\ra \Omega^j_\mr{distr}(X)$ of Morse complex with real coefficients into distributional forms. It maps a critical point $P$  of $F$ of coindex $j$ to the 
delta-form on the unstable manifold of $P$ (a.k.a. Lefschetz thimble),
\begin{equation}\label{i}
i\colon [P]\mapsto \delta_{\mr{Unstab}_P}.
\end{equation} 
\item Projection $p\colon \Omega^{j}_\mr{distr}(X)\ra MC^j(F,\RR)$ is defined by 
\begin{equation}\label{p}
p\colon \omega\mapsto \sum_{P\in \mr{Crit}(F)} \left(\int_{\mr{Stab}_P} \omega \right) \cdot[P],
\end{equation}
where $\mr{Stab}_P\subset X$ is the stable manifold of $P$. Linear operator $p$ is densely defined and satisfies $p\circ i=\mr{id}_{MC(F)}$.
\item The chain homotopy $K\colon \Omega^j_\mr{distr}(X)\ra \Omega^{j-1}_\mr{distr}(X)$ is defined as 
\begin{equation} \label{K}
K= \int_{0}^\infty U(t,dt) \quad =-\int_0^\infty dt\; \iota_v \; e^{-t \mc{L}_v}\;\;\colon \quad \Omega^j_\mr{distr}(X) \ra \Omega^{j-1}_\mr{distr}(X).
\end{equation}
It is also a densely defined operator, satisfying the chain homotopy property 
\begin{equation}\label{dK+Kd=id-ip}
dK+K d=i\, p-\mr{id}.
\end{equation}
\end{itemize}

Here is a sketch of proof of (\ref{dK+Kd=id-ip}): writing $U=e^{-(d_t+[d_X,-])(t\iota_v)}$ we see that $U$ satisfies 
\begin{equation}\label{(d+Q)U=0}
(d_t+[d_X,-])U=0.
\end{equation} 
Thus, $[d_X,K]=-\int_0^\infty [d_X,U]=\int_0^\infty d_t U=U(+\infty)-U(0)=U(+\infty)-\mr{id}$. 
We still need to show that
\begin{equation}\label{U(infty)=ip}
U(+\infty)=i\circ p.
\end{equation} 
For that, note that 
the $0$-form component  along $(0,+\infty)$ of the operator $U(t,dt)$ is an integral operator with distributional kernel $\delta_{Y_t}$, where 
\begin{equation}
Y_t=\{(x,y)\in X\times X\;|\; x=\mr{Flow}_t(v) (y)\}\quad \subset X\times X
\end{equation}
is the submanifold consisting of pairs of points $(x,y)$, such that the gradient trajectory of starting at $y$ ends up at $x$ after time $t$. For $t\ra+\infty$, we are looking at pairs of points connected by a gradient trajectory which takes a very long time $t$ -- thus, it is a ``broken trajectory,'' passing near (or at $t=+\infty$, through) a critical point. Therefore, $Y_{+\infty}=\cup_P \mr{Unstab}_P\times\mr{Stab}_P$. Thus, $U(+\infty)$ is the integral operator with distributional kernel $\sum_P \delta_{\mr{Unstab}_P}(x)\delta_{\mr{Stab}_P}(y)$. By inspection, it coincides with $i\circ p$.

We note that the chain homotopy operator $K$ can also be seen as an integral operator with distributional kernel $\delta_{\widetilde{Y}}$ where 
\begin{equation}\label{Y tilde}
\widetilde{Y}=\bigcup_{t>0}Y_t\quad \subset X\times X.
\end{equation}
From this standpoint the relation (\ref{dK+Kd=id-ip}) is immediate from the relation at the level of singular supports of integral kernels, $\dd \widetilde{Y}=Y_{+\infty}-\mr{Diag}$, where $\mr{Diag}=\{(x,x)\}\subset X\times X$ is the diagonal.

\textbf{Regularized Morse contraction.}

For the following we need to introduce regularization data -- a family, parametrized by $\epsilon>0$, of closed forms $\rho_\epsilon\in \Omega^{\dim X}(X\times X)$ supported in $\epsilon$-neighborhood $N_\epsilon$ of the diagonal $\mr{Diag}\subset X\times X$, such that their cohomology class is Poincar\'e dual to the homology class $[\mr{Diag}]$. 
Denote $R_\epsilon\colon \Omega^\bt_\mr{distr}(X)\ra \Omega^\bt(X)$ the ``smearing operator,'' i.e. the linear operator from distributional to smooth forms defined by the integral kernel $\rho_\epsilon$. Then for a chain $Z$, one has the regularized (smeared)
 delta-form $$\delta^\epsilon_Z\colon= R_\epsilon(\delta_Z)$$ 
 supported in the $\epsilon$-neighborhood of $Z$. If chains $Z_1$ and $Z_2$ intersect transversally and the intersection is a $0$-chain, then for $\epsilon$ sufficiently small one has 
\begin{equation} \label{epsilon-regularized intersection}
 \int_X \delta^\epsilon_{Z_1}\wedge \delta^\epsilon_{Z_2}=\int_X \delta^\epsilon_{Z_1}\wedge \delta_{Z_2}=\#(Z_1\cap Z_2)
\end{equation}
-- the oriented count of intersection points.

One can also construct (Lemma 2 in \cite[Section 6.5]{KS}) a family of operators $h_\epsilon\colon \Omega^\bt(X)\ra \Omega^{\bt-1}(X)$ with integral kernel supported in $N_\epsilon$, such that ${dh_\epsilon+h_\epsilon d} = R_\epsilon|_{\Omega(X)}-\mr{id}$.

 One has a deformation retraction of de Rham complex $(\Omega(X),d)$ onto the Morse chain complex with real coefficients $(MC(F,\RR),d_\mr{Morse})$ given by the following triple of maps. 
\begin{itemize}
\item The regularized inclusion $i^\epsilon\colon MC^j(F,\RR)\ra \Omega^j(X)$ maps a critical point $P$  of $F$ of coindex $j$ to the 
\emph{$\epsilon$-smeared} delta-form on the unstable manifold of $P$,
\begin{equation}\label{i_epsilon}
i^\epsilon\colon 
[P]  \mapsto  {\delta_{\mr{Unstab}_P}^\epsilon}.
\end{equation} 
\item 
Projection $p\colon \Omega^{j}(X)\ra MC^j(F,\RR)$ is defined again by the formula (\ref{p}), restricted to smooth forms.
Then, for sufficiently small $\epsilon$, one has $p\circ i^\epsilon=\mr{id}_{MC(F)}$.
\item The regularized chain homotopy $K^\epsilon\colon \Omega^j(X)\ra \Omega^{j-1}(X)$ is defined as 
\begin{equation} \label{K epsilon}
K^\epsilon = R_\epsilon \circ K + h_\epsilon,
\end{equation}
where $K$ is defined by (\ref{K}) restricted to smooth forms.
It satisfies the property 
\begin{equation}\label{dK+Kd=id-ip regularized}
dK^\epsilon+K^\epsilon d=i^\epsilon\, p-\mr{id}.
\end{equation}
\end{itemize}
We will call the triple of maps 
\begin{equation}
K^\epsilon\;\; \rotatebox[origin=c]{90}{$\curvearrowright$}
\quad  (\Omega(X),\;d) \quad \stackrel{i^\epsilon}{\underset{p}{\leftrightarrows}}\quad  (MC(F),\; d_\mr{Morse})
\end{equation}
defined above the \emph{(regularized) Morse contraction} associated with a Morse function $F$.

\subsection{Example: deformation of the Morse differential by cycles}
\label{example: Novikov}
Consider the category $\FF$ with two objects $F_1,F_2$. 
Let $\{P_i\}$ be the critical points of $F=F_1-F_2$. Fix a collection of cycles $\{C_\alpha\}$ on $X$ (seen as elements of $\Mor(F_2,F_2)$). Consider the generating function
for higher compositions
\begin{equation} \label{N=2 compositions}
m\colon \Mor(F_1,F_2)\otimes \Mor(F_2,F_2)^{\otimes k} \ra \Mor(F_1,F_2) 
\end{equation} 
given by 
\begin{equation}\label{m_ij}
m_i^j(T)=\sum_{k\geq 1}\sum_{\alpha_1,\ldots,\alpha_k} 
\#\M(P_i,C_{\alpha_1},\ldots,C_{\alpha_k},P_j)\;T_{\alpha_1}\cdots T_{\alpha_k}, 
\end{equation}
where $T_\alpha$ are generating parameters, of parity opposite to parity of codimension of $C_\alpha$ (one can also assign a $\ZZ$-grading by $|T_\alpha|=1-\mr{codim}\, C_\alpha$). Thus, coefficients of monomials in $T$ in (\ref{m_ij}) are the counts of gradient trajectories of $F$ from $P_i$ to $P_j$ passing through given cycles.
 The generating function (\ref{m_ij}) is a matrix element of an endomorphism $m(T)$ of $\mathbb{Q}\otimes \Mor(F_1,F_2)[[T_\alpha]]$ (see Remark \ref{rem: Novikov repeated cycle regularization} below for the explanation of the $\mathbb{Q}$ factor). The $A_\infty$ relations (\ref{A_infty rel}) imply the Maurer-Cartan equation
\begin{equation}\label{d+m(T) squares to zero}
[d_\mr{Morse},m(T)]+m(T)^2=0, 
\end{equation}
cf. \cite{Lysov}. Equivalently, $d_\mr{Morse}+m(T)$ is a differential. 

We remark that the ``generating function'' $m(T)$ forgets a part of the data of compositions (\ref{N=2 compositions}): it implicitly involves a symmetrization over cycles (forgets the order in which the gradient trajectory intersects them) -- since the generating parameters $T_\alpha$ are considered as supercommutative.

\begin{remark} \label{rem: Novikov repeated cycle regularization}
The transversality assumption (Assumption \ref{assump: transversality}) fails when several repetitions of the same cycle occur as arguments in (\ref{m_ij}). We resolve the failure of transverality by slightly displacing copies of the cycle and symmetrizing over which copy of the cycle receives which displacement. 
Example: consider the case of a single cycle $C$ of codimension $1$ and consider the term $\#\M(P_i,C,C,P_j)\; T^2$ in (\ref{m_ij}). Displacing the two copies of $C$ to $C',C''$ and symmetrizing, we have $\frac12(\#\M(P_i,C',C'',P_j)+\#\M(P_i,C'',C',P_j))\; T^2$ -- so, we are counting gradient trajectories from $P_i$ to $P_j$ passing first through $C'$, then through $C''$ (with weight $\frac{T^2}{2}$) and passing first through $C''$, then through $C'$ (with weight $\frac{T^2}{2}$). For small displacements, this is the same as 
$\frac12 \sum_\gamma (\#(\gamma\cap C))^2$, where the sum is over  gradient trajectories $\gamma$ from $P_i$ to $P_j$.
We note that this regularization of repeating cycles results in the appearance of rational (as opposed to integer) coefficients in $m(T)$.
\end{remark}

\begin{example}

\label{example: circle with a cycle}
Here is a toy example: let $X=S^1$ with the height function $F$ coming from embedding $X$ as a unit circle in $\RR^2$, with a critical point $P$ of index $1$ and a critical point $Q$ of index $0$. Let $C$ be a $0$-cycle -- some point on $S^1\backslash\{P,Q\}$, and let $T$ be the corresponding generating parameter (commuting, of degree $|T|=0$).
\begin{figure}[H]
$$ \vcenter{\hbox{ \includegraphics[scale=0.8]{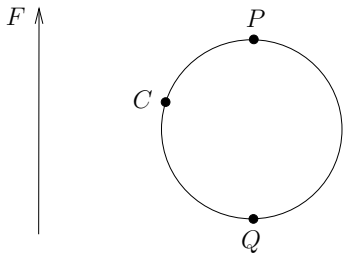} }}$$
\caption{Circle with a cycle $C$.}
\end{figure}
Then the Morse differential $d_\mr{Morse}=0$ but the deformed Morse differential is
$$ 
d_\mr{Morse}+m(T)\colon  \left\{ 
\begin{array}{ccc}
[P] &\mapsto& (e^T-1) [Q] \\
{[Q]} &\mapsto& 0 
\end{array}
\right.
$$
Here the deformation comes from
$$ m^Q_P(T)= \#\M(P,C,Q)\; T + \#\M(P,C,C,Q)\;  T^2 + \cdots = T+\frac12 T^2 + \cdots = e^T-1.$$
For each displacement of the copies of $C$, there are either one or zero gradient trajectories from $P$ to $Q$ passing the copies in the correct order. After symmetrization over displacements, we get the coefficients $\frac{1}{k!}$.

\end{example}

\textbf{Approach through homological perturbation lemma.}
The relation (\ref{d+m(T) squares to zero}) can be proven by considering the complex of differential forms on $X$ as a bicomplex, with the first differential being
de Rham $d_1=d$ and the second being multiplication by the differential form 
$$d_2=\sum T_\alpha \delta_{C_\alpha}^\epsilon,$$  
where $\delta^\epsilon_{C_\alpha}$ is the smeared delta-form of a cycle, as in Section \ref{sec: Morse contraction}.
When $C_\alpha$ are of codimension one, such differential
is known as Novikov differential $d_\mr{Novikov}=d+\omega$; what we are using here is its obvious generalization.
The operator $ m(T)$ is just a result of induction of the second differential on a subcomplex (w.r.t. $d_1$) obtained by the homological perturbation lemma,\footnote{See \cite{Gugenheim,LambeStasheff}. For a review, see
e.g. \cite{Crainic}. Also, see \cite{Losev} for the topological quantum mechanics perspective.} using Morse contraction.  The deformed de Rham complex $(\Omega^\bt(X)[[T_\alpha]],\; d_1+d_2)$ is quasi-isomorphic to the deformed Morse complex $(MC^\bt(F,\RR)[[T_\alpha]],\; d_\mr{Morse}+m(T))$ with 
\begin{equation} \label{m(T) via HPT}
m(T)= pd_2 i^\epsilon+pd_2 K^\epsilon d_2 i^\epsilon +p d_2 K^\epsilon d_2 K^\epsilon d_2 i^\epsilon+\cdots 
\end{equation}
Here $(i^\epsilon,p,K^\epsilon)$ are the maps (\ref{i}), (\ref{p}), (\ref{K}) above. 
We are assuming $\epsilon$ to be sufficiently small.
Indeed, expanding (\ref{m(T) via HPT}) we have
\begin{multline}
m(T)[P_i] = \\
\hspace{-1cm}
\lim_{\epsilon\ra 0}\sum_j\sum_{k\geq 1}\sum_{\alpha_1,\ldots,\alpha_k}   \left( \int_{\mr{Stab}_{P_j}} \delta^\epsilon_{C_{\alpha_k}} K^\epsilon(\delta^\epsilon_{C_{\alpha_{k-1}}}\cdots \delta^\epsilon_{C_{\alpha_{2}}} K^\epsilon (\delta^\epsilon_{C_{\alpha_1}} \delta^\epsilon_{\mr{Unstab}_{P_i}} )\cdots) \right)T_{\alpha_1}\cdots T_{\alpha_k} \cdot  [P_j]
\\
=\sum_j \sum_{k\geq 1} \sum_{\alpha_1,\ldots,\alpha_k} \#\M(P_i,C_{\alpha_1},\ldots, C_{\alpha_k},P_j) T_{\alpha_1}\cdots T_{\alpha_k} \cdot  [P_j].
\end{multline}
 The last equality is obvious except for the more subtle case when the sequence $C_{\alpha_1},\ldots, C_{\alpha_k}$ has repeating cycles. We expect that in this case also the contribution arising from $\epsilon$-regularization agrees with the enumerative prescription of Remark \ref{rem: Novikov repeated cycle regularization}.\footnote{ 
E.g. in the Example \ref{example: circle with a cycle}, this is a straightforward check: choose a regularization data in the sense of Section \ref{sec: Morse contraction} -- a closed 1-form $\rho_\epsilon=f_\epsilon(\theta,\theta')d\theta+g_\epsilon(\theta,\theta')d\theta'$  on $S^1\times S^1$ supported in the $\epsilon$-neighborhood of the diagonal satisfying $\int f_\epsilon d\theta=1$, $\int g_\epsilon d\theta' = -1$ ($\theta$ is the angle coordinate on $S^1$).
One proves by induction in $k=1,2,\ldots$ that $(\delta^\epsilon_C K^\epsilon)^{k-1}(\delta^\epsilon_C\delta^\epsilon_{\mr{Unstab}_P})=\frac{1}{k!}d\theta\, \frac{d}{d\theta}
(\int^\theta d\theta' f_\epsilon(\theta',\theta_0))^{k}$ where $\theta_0$ is the position of the 0-cycle $C$. This implies $\int_{\mr{Stab}_Q} (\delta^\epsilon_C K^\epsilon)^{k-1}(\delta^\epsilon_C\delta^\epsilon_{\mr{Unstab}_P}) = \frac{1}{k!}$.
}
Thus, with this caveat, (\ref{m(T) via HPT}) agrees with  the enumerative definition (\ref{m_ij}).
The fact that $d_\mr{Morse}+m(T)$ squares to zero is then a direct  consequence of the homological perturbation lemma. 


\subsection{Intersection formula for the moduli space of Morse trees. Transversality assumption.}\label{ss: transversality}
Consider the moduli space $\M$ of Morse trees appearing in the definition of the composition map (\ref{mu}). There is a natural injective map from it to $X^m$, given by recording 
the positions in $X$ of trivalent vertices of a Morse tree and the positions of intersections of gradient trajectories with chains $Z_{\alpha,a}$. Here $m=r-2+\sum_{i}k_i$ is the total number of internal (2- and 3-valent) vertices in relevant trees. 
The image of this injection -- and hence a model for $\M$ -- is  given in terms of intersections of chains in $X^m$:
\begin{multline}\label{intersection formula for M}
\M(\{Z_{\alpha,a_1}\}_{\alpha=1}^{k_1};P_{a_1 a_2};\{Z_{\alpha,a_2}\}_{\alpha=1}^{k_2};P_{a_2a_3};\cdots;P_{a_{r-1}a_r};\{Z_{\alpha,a_r}\}_{\alpha=1}^{k_r};P_{a_1a_r})
\simeq\\
\hspace{-1cm}
\sum_{\mr{trees}\,T} \left(\left(\bigcap_{e\in E} \pi^{-1}_e \widetilde{Y}_e  \right) \cap\left(\bigcap_{q\in V_2} \pi^{-1}_q(Z_q)\right)\cap \left(\bigcap_{i=1}^{r-1} \pi_{\mr{leaf}\,i}^{-1}(\mr{Unstab}_{P_{a_i a_{i+1}}})\right) \cap \pi^{-1}_\mr{root}(\mr{Stab}_{P_{a_1a_r}}) \right).
\end{multline}
Here:
\begin{itemize}
\item The sum is over combinatorial types of rooted trees $T$ with $r-1$ leaves, with edges carrying 
bi-color
$(ab)$ as above, where we encounter $k_i$ 2-valent vertices of color $a_i$ along the  path from leaf $(a_{i-1},a_i)$ to leaf $(a_i,a_{i+1})$ (where leaf $(a_r, a_1)$  is the root). Vertices in $T$ are assumed to be at most 3-valent.
\item $E$ is the set of
internal edges (not adjacent to 1-valent vertices).
$V=V_2\cup V_3$ is the set of vertices of valence $\neq 1$; $V_k$ is the set of $k$-valent vertices.
\item For an edge $e$ of bi-color $(a,b)$,
$\til{Y}_e$ is the chain (\ref{Y tilde}) in $X\times X$ associated with the Morse function $F_a-F_b$.
\item Map $\pi_{q}\colon X^V \ra X$ is the projection to $q$-th component. For $e$ an edge connecting vertices $q_1$ and $q_2$, the map $\pi_{e}\colon  X^V \ra X\times X$  is the projection selecting $q_1$-th and $q_2$-th copies of $X$. $\pi_{\mr{leaf}\,i}\colon X^V\ra X$ selects the component corresponding to the vertex connected by an edge to the $(a_i,a_{i+1})$-leaf. Likewise, $\pi_\mr{root}\colon X^V\ra X$ selects the component corresponding to the vertex connected by an edge to the root.
\end{itemize}

The key transversality assumption on the sequence of critical points and chains, under which the composition (\ref{mu}) is well-defined is the following.
\begin{assumption}\label{assump: transversality}\emph{
We assume that the sequence of input critical points $P_{ab}$ and chains $Z_{\alpha,a}$ in (\ref{intersection formula for M}) is such that for each tree $T$ the intersection in the right hand side of (\ref{intersection formula for M}) is transversal.}
\end{assumption}

Under this assumption, $\M$ is a smooth integral chain in $X^m$ of dimension
\begin{multline*}
\dim\M=\\
=(\mr{coind}(P_{a_1a_r})-1)-\sum_{i=1}^{r-1} (\mr{coind}(P_{a_ia_{i+1}})-1)-\sum_{\alpha,k} (\mr{codim}(Z_{\alpha,k})-1) -1. 
\end{multline*} 

\begin{example} Here is an example of a non-transversal composition: $N=2$, $X$ is a 2-sphere realized as a unit sphere in $\RR^3$, with $F_1-F_2$  the $z$-coordinate. The critical points are the North pole $\mathsf{N}$ and the South pole $\mathsf{S}$. Let the 1-cycle $C$ be the equator $z=0$ and let  $C'$ be a 0-cycle -- a point somewhere in $S^2$. Consider the composition $m(\mathsf{N},C,C')$:
\begin{enumerate}
\item\label{transversality example 1} If the point $C'$  is in the lower hemisphere (but not at $\mathsf{S}$), there is a unique gradient trajectory connecting $\mathsf{N}$ to $\mathsf{S}$ and passing through $C$ and $C'$, so $m(\mathsf{N},C,C')=\mathsf{S}$.
\begin{figure}[H]
$$ \vcenter{\hbox{ \includegraphics[scale=0.7]{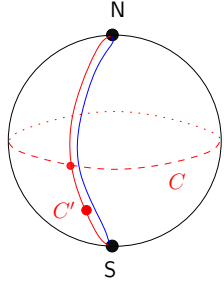} }}$$
\caption{Gradient trajectory passing through cycles $C$ and $C'$.}
\end{figure}
\item\label{transversality example 2} If the point $C'$ is in the upper hemisphere (but not at $\mathsf{N}$), then the gradient trajectory passes the cycles in wrong order, so $m(\mathsf{N},C,C')=0$.
\item \label{transversality example 3} If $C'$ is on the equator, the transversality assumption fails. 
\item\label{transversality example 4} If $C'$ coincides with one of the critical points, $\mathsf{N}$ or $\mathsf{S}$, the transversality assumption fails also. The moduli space of Morse trajectories from $\mathsf{N}$ to $\mathsf{S}$ passing through $C$ and $C'$ is one-dimensional, while the expected dimension is zero.
\end{enumerate}
So, the transversality assumption holds in situations (\ref{transversality example 1}), (\ref{transversality example 2}) and fails in situations (\ref{transversality example 3}), (\ref{transversality example 4}).
\end{example}

\subsection{An integral formula for the coefficients of composition maps} 

As a consequence of (\ref{intersection formula for M}), under the transversality assumption, one has
the following integral formula for the number of Morse trees appearing as coefficients in (\ref{mu}):
\begin{multline}\label{integral formula for |M|}
\#\M
(\{Z_{\alpha,a_1}\}_{\alpha=1}^{k_1};P_{a_1 a_2};\{Z_{\alpha,a_2}\}_{\alpha=1}^{k_2};P_{a_2a_3};\cdots;P_{a_{r-1}a_r};\{Z_{\alpha,a_r}\}_{\alpha=1}^{k_r};P_{a_1a_r})=
\\
=\sum_{\mr{trees}\, T}\int_{\RR_+^{E}} \int_{X^V}  \prod_{e\in E} \til\pi_{e}^* \eta_e  \prod_{q\in V_2} \pi_q^* (\delta_{Z_q} ) \prod_{i=1}^{r-1} \pi^*_{\mr{leaf}\,i} \left(\delta_{\mr{Unstab}_{P_{a_i a_{i+1}}}} \right)\cdot \pi^*_\mr{root}\left(\delta_{\mr{Stab}_{P_{a_1 a_r}}} \right).
\end{multline}
The notations are as in (\ref{intersection formula for M}). Additionally,
\begin{itemize}
\item Distributional form 
\begin{equation}
\eta_e=(1-dt\,\iota_v)\, \delta(Y_t)\quad  \in\;\; \Omega^{\dim X}(\RR_+\times X\times X)
\end{equation} 
is the integral kernel of the operator-valued differential form $U$ defined in (\ref{U}), associated to the gradient vector field of $F_a-F_b$ where $(a,b)$ is the bi-color of the edge $e$.
\item 
For $e$ an edge connecting vertices $q_1$ and $q_2$, the map $\til\pi_{e}\colon  \RR_+^E\times X^V \ra \RR_+\times  X\times X$  is the projection selecting $q_1$-th and $q_2$-th copies of $X$ and $e$-th copy of $\RR_+$. 
\end{itemize}

Evaluating the integral over $\RR_+^E$, we can write (\ref{integral formula for |M|}) as
\begin{equation}\label{integral formula for |M| with times integrated out}
\#\M
=\sum_{\mr{trees}\, T} \int_{X^V}  \prod_{e\in E} \pi_{e}^* \delta_{\til{Y}_e}  \prod_{q\in V_2} \pi_q^* (\delta_{Z_q} ) \prod_{a=1}^{N-1} \pi^*_{\mr{leaf}\,a} \left(\delta_{\mr{Unstab}_{P_{a_ia_{i+1}}}} \right)\cdot \pi^*_\mr{root}\left(\delta_{\mr{Stab}_{P_{a_1 a_r}}} \right).
\end{equation}
Here $\delta_{\til Y_e}$ is the integral kernel of the non-regularized Morse homotopy (\ref{K}). The product of distributional forms under the integral in (\ref{integral formula for |M| with times integrated out}) is well-defined, since the wavefronts are transversal, by Assumption \ref{assump: transversality}.

\subsection{Aside: other possible models for $\Mor(F_a,F_a)$}
\label{s: models for End}
\begin{enumerate}[(i)]
\item One can consider a variant $\FF_\Omega$ of the $A_\infty$ category $\mathbb{F}$ where the space of endomorphisms of an object $\Mor(F_a,F_a)$ is the de Rham dg algebra $\Omega^\bt(X)$. One defines composition maps by (\ref{mu}), with chains $Z_{\alpha,a}$ replaced by differential forms $\omega_{\alpha,a}\in \Omega^\bt(X)$, with coefficients $\#\M$ defined by the integral formula (\ref{integral formula for |M|}) (instead of the enumerative definition of Section \ref{ss: definition of F}),  with $\delta_{Z_q}$ replaced by smooth forms $\omega_q$. The benefit here is that, assuming the functions $F_a$ are in general position, the resulting $A_\infty$ category $\FF_\Omega$ has always well-defined compositions. The drawback however is that integral structure is lost -- morphism spaces and compositions are cochain complexes over $\RR$, not $\ZZ$. Also, the enumerative flavor of compositions is lost in this model.
We remark also that restricting compositions in  $\mathbb{F}_\Omega$ to $\epsilon$-smeared delta-forms of chains, 
one obtains compositions in $\mathbb{F}$.
\item Another option is to define the space of endomorphisms of an object as singular \emph{cochains} instead of chains, $\Mor(F_a,F_a)=C^\bt_\mr{sing}(X,\mathbb{Z})$, with dg algebra structure given by the coboundary operator and the cup product.\footnote{
P.M. thanks Stephan Stolz for this remark.
} These operations are always well-defined, without extra transversality requirement. However, when defining more general composition maps, one would still need to impose a transversality requirement. So, one still gets an $A_\infty$ category with partially defined compositions.
\end{enumerate}

\section{Proof of $A_\infty$ relations via second quantization}\label{sec 3}

\subsection{Proof by 
homotopy transfer}
Consider the differential graded 
algebra obtained by taking a tensor product of the supercommutative  de Rham dg algebra of differential forms on $X$ and the associative algebra $M_N$ of $N\times N$ matrices.
As a vector space it is
$$
V=\bigoplus_{a,b=1}^{N} \Omega_{ab} ,
$$ 
where each space $\Omega_{ab}$ is isomorphic to the space of differential forms on $X$.
Now, for $a \neq b$ consider the (regularized) Morse contraction of the space $\Omega_{ab}$ to Morse chains of the function $F_a-F_b$ with homotopy (\ref{K epsilon}).
Thus, we have a contraction of $V$ onto the complex
$$ \F=\bigoplus_{a,b} \F_{ab} \qquad \mbox{where}\quad 
\F_{ab}=\left\{
\begin{array}{ccc}
MC(F_a-F_b,\RR) & \mbox{if} & a\neq b \\
\Omega_{aa} & \mbox{if} & a=b
\end{array}
 \right.
$$

We have again a triple of maps -- inclusion, projection and chain homotopy:
\begin{equation}\label{Morse contraction big}
\ul{K} \;\; \rotatebox[origin=c]{90}{$\curvearrowright$}\quad V \qquad \stackrel{\ul{i}}{\underset{\ul{p}}{\leftrightarrows}}\qquad  \F
\end{equation}
which are given by regularized Morse contraction maps in $(a,b)$-sectors for $a\neq b$ and are trivial ($i=p=\mr{id}$, $K=0$) in $(a,a)$-sectors.

Homotopy transfer of a dg algebra 
structure on $V$ to $\MM$, via the Kontsevich-Soibelman sum-over-trees formula \cite{KS}, yields an $A_\infty$ algebra structure on $\F$. The induced higher multiplications $\mu\colon \F^{\otimes n}\ra \F$
are given as follows: 
\begin{equation} \label{KS sum-over-trees formula}
\mu(X_1,\ldots,X_n)=\sum_T \mu_T(X_1,\ldots,X_n) .
\end{equation}
Here:
\begin{itemize}
\item $T$ runs over (combinatorial) binary rooted trees with $n$ leaves.
\item $\mu_T\colon \F^{\otimes n}\ra \F$ is a multilinear map obtained by decorating the $i$-th leaf of the tree by  $\ui(X_i)$, calculating the product  in $V$ at each internal vertex, applying the chain homotopy $K$ at each internal edge, and finally applying the projection $\up$ at the root.
For example:
$$ \mu_{\vcenter{\hbox{ \includegraphics[scale=0.2]{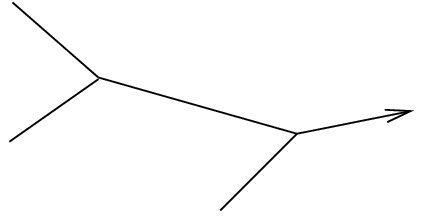} }}}(X_1,X_2,X_3)=
\vcenter{\hbox{ \includegraphics[scale=0.7]{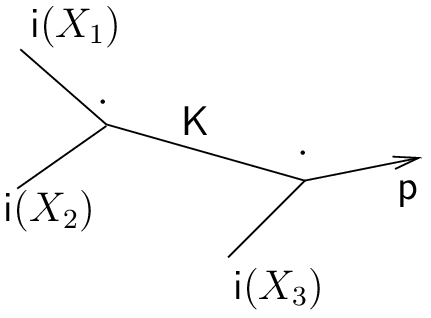} }}
=\up(\uK(\ui(X_1)\cdot \ui(X_2))\cdot \ui(X_3)).
$$
\end{itemize}

The relation between this induced $A_\infty$ algebra structure on $\F$ and the $A_\infty$ category $\FF$ is as follows.
Assume we are given a composable sequence of morphisms 
\begin{equation}\label{composable sequence}
F_{a_1}\xra{x_1}F_{a_2}\xra{x_2}\cdots \xra{x_{n}}F_{a_{n+1}},
\end{equation}
where $x_i$  is a critical point (or more generally a Morse chain) of $F_{a_i}-F_{a_{i+1}}$ if $a_i\neq a_{i+1}$ and a chain if $a_i=a_{i+1}$. Then we have
\begin{equation}\label{A_infty cat vs A_infty alg}
m(x_1,\ldots,x_n)=\mu(X_1,\ldots,X_n).
\end{equation}
Here on the left we have a composition in the category $\FF$, on the right we have a higher multiplication in the algebra $\MM$. The inputs on the right are:  $X_i=x_i$ if $a_i\neq a_{i+1}$; 
  $X_i=\delta^\epsilon_{x_i}
 $
 -- the $\epsilon$-smeared delta-form of the chain $x_i$ -- if $a_i=a_{i+1}$. 
We understand each input $X_i$ as an element of $\MM_{a_i a_{i+1}}$. The equality (\ref{A_infty cat vs A_infty alg}) holds for the regulator $\epsilon$ sufficiently small\footnote{One cannot choose a ``sufficiently small'' $\epsilon$ uniformly for all compositions in $\FF$: the bound for $\epsilon$ depends on the input chains in (\ref{A_infty cat vs A_infty alg}) -- ultimately because the bound on $\epsilon$ in (\ref{epsilon-regularized intersection}) depends on the input chains.} and follows from comparison of (\ref{KS sum-over-trees formula}) with (\ref{integral formula for |M| with times integrated out}) and (\ref{mu}).\footnote{For completeness we remark that there is also the special case when the first and last object in (\ref{composable sequence}) coincide, $a_1=a_{n+1}$. In that case, $m(x_1,\ldots,x_n)$ is a chain (see the discussion before Theorem \ref{thm 1} in Section \ref{ss: definition of F}), while $\mu(X_1,\ldots,X_n)$ is an $\epsilon$-smeared delta-form of that chain.
}
\begin{remark}
 In (\ref{A_infty cat vs A_infty alg}), the contribution of a binary rooted tree $T$ from Kontsevich-Soibelman formula (\ref{KS sum-over-trees formula}) corresponds to the contribution of a rooted tree $T'$ in (\ref{integral formula for |M| with times integrated out}) with 2-valent vertices allowed, where $T'$ is obtained from $T$ by removing the 1-valent vertices (and the adjacent edges) in $T$ colored by inputs in $\F_{a_i a_{i+1}}$ with $a_i=a_{i+1}$. Each such removal creates a 2-valent vertex in $T'$.
 
 \begin{figure}[H]
$$ \vcenter{\hbox{ \includegraphics[scale=1]{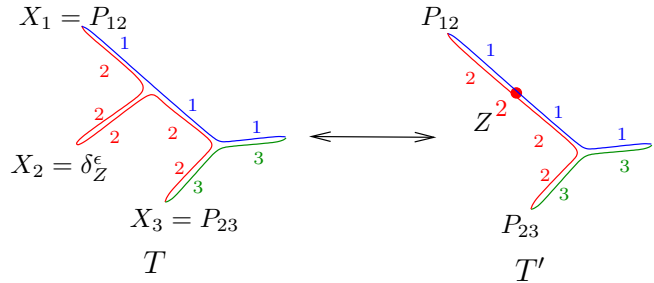} }}$$
\caption{Example: a non-binary tree contributing to (\ref{integral formula for |M| with times integrated out}) arising from a binary tree contributing to Kontsevich-Soibelman homotopy transfer  formula by amputating an $(aa)$-leaf ($a=2$ in the picture). We are indicating explicitly the bi-colors of edges. 
}
\end{figure}
\end{remark}


By Kontsevich-Soibleman's general result, the induced higher multiplications $\mu$ on $\MM$ satisfy the $A_\infty$ algebra relations:
\begin{equation}
\sum_{r\geq 0,\,s\geq 1,\, r+s\leq n} \pm \mu(X_1,\ldots,X_r,\mu(X_{r+1},\ldots, X_{r+s}),X_{r+s+1},\ldots, X_n) = 0
\end{equation}
for any inputs $X_1,\ldots,X_n\in \MM$.
By (\ref{A_infty cat vs A_infty alg}) this implies the $A_\infty$ category relations (\ref{A_infty rel}) for composition maps in $\mathbb{F}$.




\subsection{Approach via effective action for the $BF$ theory
} The  method of proof we used above in physics admits an interpretation via 
 ``second quantization.'' Namely, we consider the $BF$ theory in Batalin-Vilkovisky formalism,\footnote{
Non-abelian $BF$ theory in BV formalism, as a  theory where fields are nonhomogeneous differential forms, was described in \cite{CR}.
 }
defined by the action
\begin{equation}\label{BF}
S= \int_X \langle B , dA+\frac12 [A,A] \rangle,
\end{equation}
where the field $A$ is a 
differential form with values in the Lie algebra  $\g=\mathfrak{gl}_N(\A)$ -- the Lie algebra of $N\times N$ matrices with entries in $\mathbb{A}$ -- the algebra of upper-triangular matrices of large size $\NN\times \NN$ (more properly, we are taking a direct limit $\NN\ra \infty$);
$B$ is a differential form valued in $\g^*$.

\begin{remark}
If we just consider $BF$ theory with coefficients in $N\times N$ matrices, then we will produce, via the effective action, an $L_\infty$ algebra structure on $\F$ -- the skew-symmetrization of the desired $A_\infty$ structure. Thus, with such coefficients we would lose a part of the data of higher multiplications. To remedy that, we introduce non-commutativity of the $(a,b)$-components of the field by letting them take values in $\A$. This has the effect of tensoring the $A_\infty$ algebra $\F$ by $\A$ and only then skew-symmetrizing. The resulting $L_\infty$ brackets on $\F\otimes \A$ remember the $A_\infty$ multiplications $\mu$ on $\F$ of arity up to $\NN-1$, see (\ref{mu via l}) below.
\end{remark}

We decompose $A$ into diagonal fields $A_Z$ (with respect to $M_N$ factor) and off-diagonal fields $A_W$. The names come from Z and W gauge bosons in the standard model.
Then we employ a tricky version of the gauge-fixing that generalizes the axial gauge
(it was first proposed by N. Nekrasov in a private communication with 
A. Losev):\footnote{
Note that the usual axial (or temporal) gauge is $A_0=0$, that is
 $\iota_{\frac{\partial}{\partial t}} A=0 $.
A feature 
of the generalization is that we take different vector fields for different components of the gauge field.
} 
\begin{equation}\label{CN gauge}
\iota_{v_{ab}} (A_W)_{ab}=0,
\end{equation}
where $v_{ab}$ is the gradient vector field of $F_a-F_b$. We remark that this gauge condition is equivalent to $(A_W)_{ab}$ being annihilated by the Morse chain homotopy (\ref{K}).


After such gauge fixing, we just integrate fields $A_W$ out and get an effective theory for $A_Z$ fields together with critical points which are the remnants of $A_W$ field. 

\textbf{More details.} 

The effective action is defined by the fiber BV integral 
\begin{equation} \label{fiber BV integral}
e^{\frac{i}{\hbar}S_\mr{eff}(A',B')} = \int_\Lag e^{\frac{i}{\hbar}S(A,B)} 
\end{equation}
over a Lagrangian $\Lag$ in the fiber of the odd-symplectic fibration
\begin{equation}\label{symp fibration}
\Omega^\bt(X,\g)[1]\oplus \Omega^\bt(X,\g^*)[\dim X-2] \quad \ra\quad \MM\otimes \A[1]\oplus \MM\otimes \A^*[\dim X-2].
\end{equation}
On the left we have the space of BV fields of $BF$ theory (its elements are pairs $(A,B)$), on the right we have our chosen space of ``slow fields'' -- we denote its elements  $(A',B')$.
The projection in (\ref{symp fibration}) is $\up\oplus \up$, built from maps of (\ref{Morse contraction big}) and extended by $\A$-linearity; it also has a section $\ui\oplus\ui$.
The Lagrangian $\Lag$ in the fiber of (\ref{symp fibration}) is determined by the gauge-fixing condition (\ref{CN gauge}), or in other words is given by $\Lag=\mr{im}\uK\oplus \mr{im}\uK$. 

Evaluating the fiber BV integral (\ref{fiber BV integral}) as a perturbed Gaussian integral yields
the effective action as a sum of Feynman diagrams:
\begin{equation}\label{sum over Feynman trees}
S_\mr{eff} = \sum_{\Gamma} \frac{1}{|\mr{Aut}(\Gamma)|} \Phi_\Gamma(A',B').
\end{equation}
Here 
\begin{itemize}
\item The sum is over binary rooted trees $\Gamma$ (Feynman graphs) up to graph isomorphism.
\item $|\mr{Aut}(\Gamma)|$ is the order of the automorphism group of the tree $\Gamma$.
\item $\Phi_\Gamma(A',B')$ is polynomial function on the space of slow fields (the r.h.s. of (\ref{symp fibration})). 
It is obtained by decorating leaves of $\Gamma$ with $\ui(A')$, the root with $\langle B',\up(\cdots) \rangle$, binary vertices  with the Lie bracket in the dg Lie algebra $\Omega^\bt(X,\g)$, internal edges  with chain homotopy $\uK$. 
\end{itemize}
The first terms in (\ref{sum over Feynman trees}) are:
\begin{multline}
S_\mr{eff}= \langle B', d_{\MM} A' \rangle + \frac12 \langle B', \up [\ui(A'),\ui(A')]  \rangle + \frac12 \langle B', \up [\uK [\ui(A'),\ui(A')],\ui(A')] \rangle + \\
+\frac12 \langle B', \up [\uK [\uK[\ui(A'),\ui(A')],\ui(A')] , \ui(A')] \rangle  
+\frac18 \langle B', [\uK[\ui(A'),\ui(A')],\uK[\ui(A'),\ui(A')]] \rangle
 +\cdots
\end{multline}

\textit{Description in terms of diagonal and off-diagonal fields.}
One can split the fields (and slow fields) into diagonal and off-diagonal components w.r.t. the $M_N$-factor:
$$ 
\begin{array}{cc}
A=A_Z+A_W,& B=B_Z+B_W, \\
A'=A_Z+A'_W, & B'=B_Z+B'_W.
\end{array}
 $$
Then the action (\ref{BF}) is
\begin{multline}\label{S_BF expanded in Z,W}
S=\int_X \underset{(a)}{\langle B_Z,dA_Z \rangle} + \underset{(b)}{\langle B_W,dA_W \rangle} + \underset{(c)}{\langle B_W,\frac12 [A_W,A_W] \rangle} +\\
+ \underset{(d)}{\langle B_W,[A_Z,A_W] \rangle} +
\underset{(e)}{\langle B_Z,\frac12[A_Z,A_Z]\rangle} +\underset{(f)}{\langle B_Z,\frac12 [A_W,A_W] \rangle}. 
\end{multline}
Fields $A_Z,B_Z$ are non-dynamical -- not integrated over in (\ref{fiber BV integral}), while $A_W,B_W$ are dynamical. In
 (\ref{S_BF expanded in Z,W}): 
\begin{itemize} 
\item  Terms (a), (e) are constants. 
\item (b) is a kinetic term for the dynamical field, generating the ``propagator'' $\uK$ assigned to edges. 
\item (c) generates a trivalent ``interaction'' vertex (two incoming and one outgoing half-edges). 
\item (d) generates a trivalent vertex with one loose (or ``external'') incoming half-edge decorated with $A_Z$. 
\item (f) generates a trivalent vertex with loose outgoing half-edge decorated with $B_Z$. 
\item Half-edges decorated by $A_W$ can be either external, decorated by $A'_W$, or internal (corresponding to the fast/dynamical part of this field component) -- connected to an internal $B_W$-half-edge, with resulting internal edge decorated by the propagator. Similarly, a $B_W$-half-edge can be either external or internal.
\end{itemize}
 \begin{figure}[H]
$$ \vcenter{\hbox{ \includegraphics[scale=0.8]{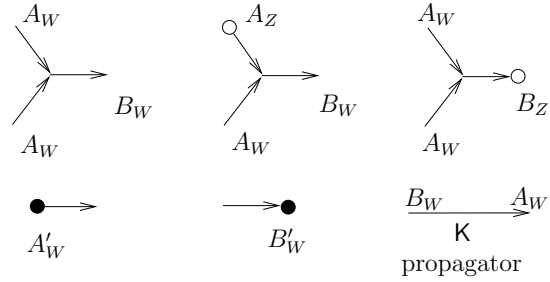} }}$$
\caption{Feynman rules (building blocks of Feynman graphs).
}
\end{figure}

 \begin{figure}[H]
$$ \vcenter{\hbox{ \includegraphics[scale=0.5]{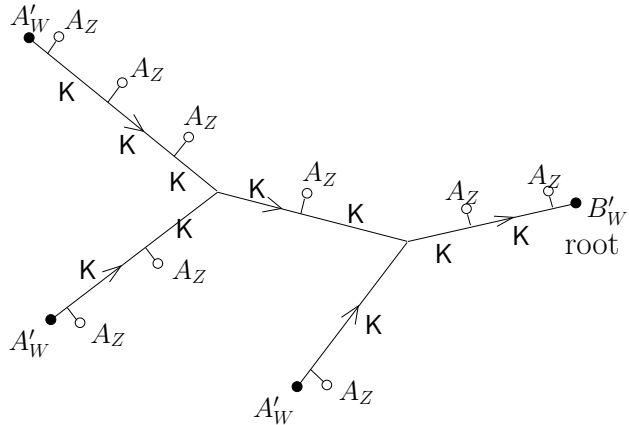} }}$$
\caption{A typical Feynman tree contributing to $S_\mr{eff}$, depicting explicitly diagonal and off-diagonal parts of fields.
}
\end{figure}

Grouping together the contributions of Feynman trees with same number of leaves, one can write $S_\mr{eff}$ in the form
\begin{equation}
S_\mr{eff}=\sum_{n\geq 1} \frac{1}{n!}\langle B' , l_n(A',\ldots,A')\rangle,
\end{equation}
where $l_n\colon \wedge^n (\F\otimes \A)\ra \F\otimes \A$
are skew-symmetric multilinear operations given by Feynman trees with $n$ leaves. Expansion in Feynman graphs (\ref{sum over Feynman trees}) yields a formula for $l_n$ as a sum over non-planar (or up-to-isomorphism) binary rooted trees, which is the Lie version of Kontsevich-Soibelman formula (\ref{KS sum-over-trees formula}).
We remark that although generally the effective action for $BF$ theory contains 1-loop diagrams, in our case they vanish, since the coefficient Lie algebra $\g$ is nilpotent.\footnote{In fact, it is quite interesting to consider non-nilpotent $\g$ -- for instance taking $\A$ to be all matrices of size $\til{N}\times \til{N}$. This would result in nontrivial 1-loop graphs in the effective action. Understanding their meaning as a 
``loop enhancement'' of the Fukaya-Morse category is a work in progress by the authors.}

We emphasize that in the logic of gauge theory, we are choosing a gauge-fixing for the integral (\ref{fiber BV integral}) over fast fields (the complement of slow fields). In BV formalism, that amounts to a choice of a Lagrangian subspace in fast fields, which -- upon converting the integral into a sum of Feynman diagrams -- corresponds to the propagator (the decoration of edges). Thus, 
our 
geometric gauge-fixing corresponds to edges being decorated by Morse chain homotopy (\ref{K}), and this in turn leads to the effective action (computed in this gauge) being related to the count of Morse trees.

The action (\ref{BF}) satisfies the BV master equation $\{S,S\}=0$ as a consequence of the quadratic relations in dg Lie algebra of $\g$-valued forms: $d^2=0$, Leibniz and Jacobi. Here $\{,\}$ is the Poisson bracket associated with the odd-symplectic form on the space of fields. Therefore, by BV-Stokes' theorem for fiber BV integrals, the effective action also satisfies the master equation 
\begin{equation}
\{S_\mr{eff},S_\mr{eff}\}=0.
\end{equation}
It is equivalent to the quadratic relations of an $L_\infty$ algebra on the operations $l_n$:
\begin{equation}
\sum_{r\geq 0, s\geq 1,\, r+s=n}\frac{1}{r!s!}\sum_{\sigma\in S_n}\pm l_{r+1}(y_{\sigma(1)},\ldots,y_{\sigma(r)},l_s(y_{\sigma(r+1)},\ldots, y_{\sigma(n)} ))=0 ,
\end{equation}
where $y_1,\ldots, y_n \in \MM\otimes \A$ are the inputs and $\sigma$ runs over permutations.
We refer to \cite{discrBF} for more details on homotopy transfer of $L_\infty$ algebras via fiber BV integrals.

\textbf{From $L_\infty$ back to $A_\infty$.}
We can recover the $A_\infty$ multiplications in $\MM$ (and thus  higher compositions in $\mathbb{F}$) from the $L_\infty$ brackets $l_n$ as follows:
\begin{equation}\label{mu via l}
l_n(X_1\otimes t_{12},\ldots, X_n\otimes t_{n,n+1})=\mu_n(X_1,\dots,X_n)\otimes t_{1,n+1},
\end{equation}
where $t_{ij}\in \A$ is the matrix with entry $1$ at position $(i,j)$ and all other entries zero. 
Here  $X_{i}$ are elements of $ \MM_{a_i b_i}$ where we require that $b_i=a_{i+1}$ (otherwise both sides of (\ref{mu via l}) vanish).  


\section{Proof of $A_\infty$ relations via first quantization
}
\label{sec 4}
\subsection{
Approach to $A_\infty$ relations via
higher topological quantum mechanics on trees: the idea}\label{ss: HTQM idea}
We give the second proof of the  $A_\infty$ relations (\ref{A_infty rel}) in the  category $\mathbb{F}$ 
 by considering higher topological quantum mechanics on trees.\footnote{See \cite{Losev,LP} for the setup of higher topological quantum mechanics (HTQM). 
Basically, HTQM is a 
closed differential form on the moduli space of metric trees which factorizes when any edge becomes infinitely long. This leads, by Stokes' theorem on the moduli space, to quadratic relations among periods of this form. This setup  is essentially a chain level, one-dimensional version of
a cohomological field theory \cite{KM}.
}

Ordinary quantum mechanics is described by the space of states of the particle and the Hamiltonian.
We will consider particles of different types  $(a,b)$ with $a \neq b $, and the space of states
will be always space of differential forms on the manifold $X$.
We will take as Hamiltonians the Lie derivatives $\mathcal{L}_{v_{ab}}$ along gradient vector fields $v_{ab}$ of differences $F_a-F_b$.
Such Hamiltonians are exact with respect to  an odd nilpotent symmetry $Q$:
$$
Q=d, 
\; \; G= \iota_{v_{ab}},\;\;
H=[ Q , G ]= \mathcal{L}_{v_{ab}}.
$$
We will consider trees equipped with a metric (lengths of edges), with vertices of valence 1, 2 and 3. 

A 1-valent 
vertex is a state that for the $(a,b)$-particle which we take to be the $\delta$-form on the Lefschetz thimble  (unstable manifold) corresponding to a critical point $P_{ab}$ of the function $F_a-F_b$.
A 2-valent 
vertex -- just an operator in the space of states -- is given by the  operator of multiplication by  the $\delta$-form on a chain $Z$.
The 3-valent 
vertex is the multiplication of differential forms together with the rule
$$
(a,b) \times (b,c)=(a,c)
$$
and zero otherwise.
We construct a differential form on the space of lengths of edges  by using the superpropagator (a.k.a. evolution operator)
\begin{equation}\label{U in HTQM}
U(t,dt)=\exp( -t H -dt G).
\end{equation}
This object appeared in our discussion of the Morse contraction above (\ref{U}).
Expanding this propagator, we get a differential form on the space of lengths of edges. This form factorizes when
the length of an edge goes to infinity.

The rough idea of the argument is as follows (the correct version is presented below). The differential form constructed in such a way is closed. Therefore we may
integrate it along the infrared boundary (when length of one of the edges equals infinity).
Using factorization, we get the desired quadratic relations between integrals over the space of lengths, which are exactly the 
 $A_\infty$ relations (\ref{A_infty rel}) we want to prove. 


\subsection{More details}
 Fix a sequence of ``colors'' $a_1,\ldots,a_r\in \{1,\ldots,N\}$ with $a_i\neq a_{i+1}$ for all $i$ and a sequence of integers $k_1,\ldots,k_r\geq 0$.
We consider the moduli space $MT_{a_1,\ldots,a_r;k_1,\ldots,k_r}$ of metric trees. 
These are rooted ribbon trees with $r-1$ incoming $1$-valent vertices (leaves) 
 and one outgoing (root). 
The sides (borders) of ribbons carry a color $a_1,\ldots,a_r$ starting with $a_1$ as we go from the root counterclockwise and changing from $a_i$ to $a_{i+1}$ at $i$-th 1-valent vertex encountered.
  The trees additionally have $k_i$ bivalent vertices of color $a_i$, distributed along the edges sharing 
  the $i$-th border. 
Thus, each edge and each 1-valent vertex carry a bi-color $(a,b)$ with $a\neq b$ while each 2-valent vertex carries a single color $a$, and can occur only on a line with bi-color $(a,b)$ or $(b,a)$ for some second color $b$.

We are assigning lengths $t_j>0$ to the internal edges (i.e. those not adjacent to a 1-valent vertex -- those edges are understood as having length $+\infty$); the limit $t_j\ra 0$ is understood as contraction of an edge. 
Thus, $MT=\bigcup \RR_+^{\#\mr{int.\, edges}}/\sim$ where the union is over combinatorial types of trees of type $(a_1,\ldots,a_r;k_1,\ldots,k_r)$ and $\sim$ is the equivalence of a tree with an edge of length zero with a tree with that edge collapsed.
\begin{figure}[H]
$$ 
 \vcenter{\hbox{ \includegraphics[scale=0.8]{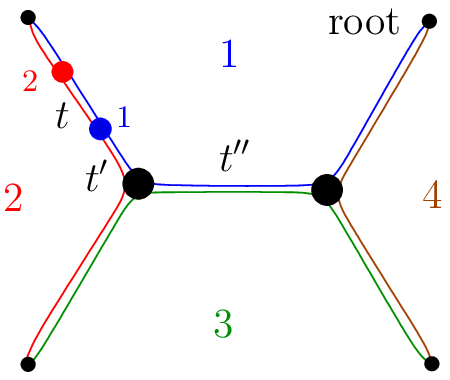} }}\qquad
  \vcenter{\hbox{ \includegraphics[scale=0.8]{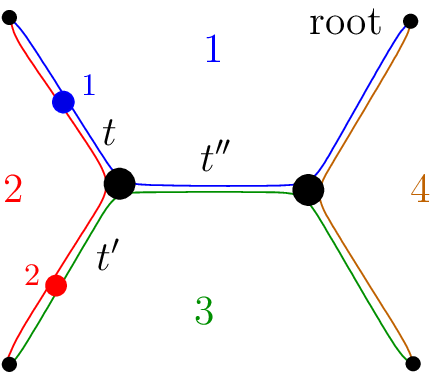} }}\qquad
 \vcenter{\hbox{ \includegraphics[scale=0.8]{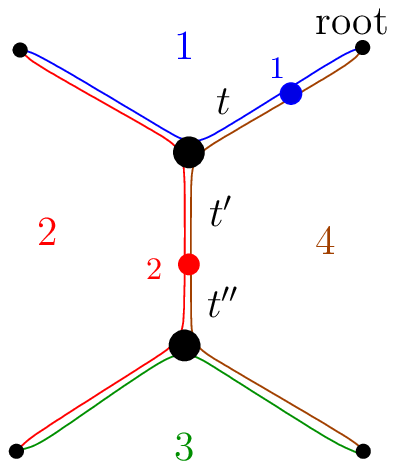} }}  
$$
\caption{Three top-dimension cells in $MT_{1,2,3,4;1,1,0,0}$.
\label{fig: MT}
}
\end{figure}

The construction above determines a differential form on the moduli space of trees (``pre-amplitude'' in the terminology of \cite{Losev}):
\begin{multline}\label{Xi}
\PA\;\in \;\Omega^\bt(MT_{a_1,\ldots,a_r;k_1,\ldots,k_r}) \otimes\\
\otimes  \mr{Hom}(
\Mor_{a_1,a_1}^{\otimes k_1}\otimes \Mor_{a_1,a_2}\otimes\cdots \otimes \Mor_{a_{r-1},a_r}\otimes \Mor_{a_r,a_r}^{\otimes k_r}\; ,\; \Mor_{a_1,a_r}
).
\end{multline}
Here we abbreviated $\Mor_{a,b}\colon= \RR\otimes \Mor(F_a,F_b)$ --  morphism spaces in $\mathbb{F}$ tensored with $\RR$. 
\begin{example}
For instance, for the middle tree in Fig. \ref{fig: MT}, we have 
$$\PA(t,dt,t',dt',t'',dt'') (Z_1, [P_{12}], Z_2, [P_{23}], [P_{34}])=\sum_{P_{14}\in \mr{Crit}(F_1-F_4)} \overline{\PA}\cdot [P_{14}] $$
with 
\begin{multline}\label{PA example}
\overline{\PA}(t,dt,t',dt',t'',dt'')= 
\int_X \delta_{\mr{Stab}_{P_{14}}}\wedge \\
\wedge U_{F_1-F_3}(t'',dt'')\left(U_{F_1-F_2}(t,dt)(\delta_{Z_1}\wedge\delta_{\mr{Unstab}_{P_{12}}} )\wedge
U_{F_2-F_3}(t',dt')(\delta_{Z_2}\wedge \delta_{\mr{Unstab}_{P_{23}}})  \right)\\
\wedge \delta_{\mr{Unstab_{P_{34}}}}.
\end{multline}
Here $\int_X \delta_{\mr{Stab}_{P_{14}}}\wedge\cdots$ is the contribution of the root (the ``out-state''). 
Notation $U_{F_a-F_b}$ refers to the superpropagator (\ref{U in HTQM}) for the gradient vector field of $F_a-F_b$.\footnote{
Note that (\ref{PA example}) is the integrand in $\int_{\RR_+^E}$  in (\ref{integral formula for |M|}) -- in the term corresponding to the combinatorial tree in the middle of Figure \ref{fig: MT} -- with the integral over $X^{V\backslash \{q\}}$ evaluated where $q$ is the vertex connected to the root by an edge (the right trivalent vertex in this example).}
\end{example}

\begin{remark}
For analytical cleanliness -- if one wants the in-states to be elements of the respective states of states (rather than limit points) --
one might want to replace delta forms $\delta_{\mr{Unstab}_{P}}$ by $\epsilon$-smeared versions $\delta_{\mr{Unstab}_{P}}^\epsilon$ (and also make the regularization $\delta_Z\ra \delta^\epsilon_Z$ at bivalent vertices) and take a limit $\epsilon\ra 0$ in the expression for $I$ in the end. 
\end{remark}

The form $\PA$ has three key properties:
\begin{enumerate}[(i)]
\item \textbf{Closedness: }
\begin{equation}\label{(d+Q)Xi=0}
(d_{MT}+Q)\;\PA=0,
\end{equation} 
where $d_{MT}$ is the de Rham differential on the moduli of trees and $Q$ acts as a derivation, acting on each morphism space in (\ref{Xi}) as the corresponding differential -- as $d_\mr{Morse}$ on $\Mor(F_a,F_b)$ with $a\neq b$ and as the boundary operator on singular chains $\Mor(F_a,F_a)$. The reason for (\ref{(d+Q)Xi=0}) is that the superpropagator $U(t,dt)$ from which $\PA$ is built is $(d_t+Q)$-closed, see (\ref{(d+Q)U=0}).
\item \textbf{Factorization on the infrared boundary:} if the length of one edge in a graph becomes $+\infty$, $\PA$ factorizes. The infrared boundary of the moduli space $MT$ has the form $\bigcup_{i<j} \dd_{ij}^{IR} MT$ where
\begin{multline}\label{IR boundary factorization}
\dd_{ij}^{IR}MT_{a_1,\ldots,a_r;k_1,\ldots,k_r}\simeq\\
\bigcup_{\tiny
\begin{array}{c}
k_i'+k_i''=k_i,\\ k_j'+k_j''=k_j
\end{array}
} MT_{a_1,\ldots,a_j;k_i',k_{i+1},\ldots,k_{j-1},k_{j}'}\times MT_{a_1,\ldots,a_i,a_j,\ldots,a_r;k_1,\ldots,k_{i-1},k_i'',k_{j}'',k_{j+1},\ldots,k_{r}}.
\end{multline}
This boundary stratum correspond to cutting a tree at the $(a_i,a_j)$-edge (equivalently, letting this edge have length $t=+\infty$); the union is over the  ways to distribute the bivalent vertices of colors $a_i$ and $a_j$ across the cut. 
Factorization of $\PA$ then means
\begin{multline}\label{factorization of Xi}
\PA_{a_1,\ldots,a_r;k_1,\ldots,k_r}|_{\dd^{IR}_{ij}}=\\
\sum_{\tiny
\begin{array}{c}
k_a'+k_a''=k_a,\\ k_b'+k_b''=k_b
\end{array}
}\big\langle \pi_1^* (\PA_{a_1,\ldots,a_j;k_i',k_{i+1},\ldots,k_{j-1},k_{j}'}) , \pi_2^* (\PA_{a_1,\ldots,a_i,a_j,\ldots,a_r;k_1,\ldots,k_{i-1},k_i'',k_{j}'',k_{j+1},\ldots,k_{r}}) \big\rangle ,
\end{multline}
where $\pi_{1,2}$ are the projections onto the two factors in the r.h.s. of (\ref{IR boundary factorization}), $\langle, \rangle$ is the wedge product of differential forms on $MT$ accompanied by  substituting the output of $\pi_1^*(\PA)$ in $\Mor_{ab}$ as input of the $\pi_2^* (\PA)$; 
the subscript of $I$ is the label of the component of $MT$ onto which $I$ is restricted.
The factorization property (\ref{factorization of Xi}) follows immediately from (\ref{U(infty)=ip}).
\item \textbf{Period:}
\begin{equation}\label{Xi period}
\int_{MT} \PA = m.
\end{equation}
The integral of $\PA$ over the moduli space $MT_{a_1,\ldots,a_r;k_1,\ldots,k_r}$ is the composition map $m$ defined in (\ref{m source, target}), (\ref{mu}). This follows by construction, in particular from (\ref{K}).
\end{enumerate}

Now, the argument for the $A_\infty$ relations is as follows. The integral of (\ref{(d+Q)Xi=0}) over the moduli space yields (using Stokes' theorem on $MT$)
\begin{equation}\label{int_bdry Xi = Q int Xi}
\int_{\dd\, MT} \PA = \pm Q \int_{MT}\PA.
\end{equation}  
The r.h.s. is $\pm Qm$ which gives terms in (\ref{A_infty rel}) where the inner $m$ acts on a single input (terms with $s=1$). The l.h.s. of (\ref{int_bdry Xi = Q int Xi}) is a sum of integrals of $\PA$ over the following types of boundary strata of $MT$:
\begin{enumerate}[(a)]
\item UV strata where a length of an edge connecting two trivalent vertices goes to zero. Contributions of these strata cancel out when we sum over combinatorial types of graphs (the contributions of two resolutions/de-contractions of a 4-valent vertex cancel out).
\item UV strata where two 2-valent vertices of different colors $a,b$ collide. They cancel out between two directions of collision.
\item UV strata where a 2-valent vertex collides with a 3-valent vertex. These cancel out between two direction a 2-valent vertex can approach a 3-valent vertex when moving on the side of a ribbon.
\item UV strata where two 2-valent vertices of the same color collide. These do not cancel and contribute to the l.h.s. of (\ref{int_bdry Xi = Q int Xi}), giving rise to the terms in $A_\infty$ relations (\ref{A_infty rel}) of the form $m(\ldots, m(Z,Z'),\ldots)$ with $Z,Z'$ two chains.
\item IR strata: an internal edge gets infinite length. By (\ref{factorization of Xi}) and (\ref{Xi period}), l.h.s. of (\ref{int_bdry Xi = Q int Xi}) in these cases gives all the terms in $A_\infty$ relations where the inner $m$ involves two or more colors.
\end{enumerate}
In summary, (\ref{int_bdry Xi = Q int Xi}) exactly gives the $A_\infty$ relations in the category $\mathbb{F}$.

\begin{example}
Consider the moduli space of metric trees with colors $a_1=1$, $a_2=2$ and with $k_1=2$, $k_2=1$. Integrating $(d_{MT}+Q)\,\PA$ over it 
yields the $A_\infty$ relation (up to signs):
\begin{multline}
0=\left(Qm+\int_{\dd\, MT_{1,2;2,1}}\PA \right)(Z_1,Z'_1,[P],Z_2)=\\
=
\Big(d_\mr{Morse}m(Z_1,Z'_1,[P],Z_2)+m(\dd Z_1,Z'_1,[P],Z_2)+m(Z_1,\dd Z'_1,[P],Z_2)\\+ m(Z_1,Z'_1,d_\mr{Morse}[P],Z_2)+m(Z_1,Z'_1,[P],\dd Z_2)\Big)\\
+\Big(m(Z_1\cap Z'_1,[P],Z_2)+m(Z_1,m(Z'_1,[P]),Z_2)+m(Z_1,Z'_1,m([P],Z_2))\\+m(m(Z_1,Z'_1,[P]),Z_2)+m(Z_1,m(Z'_1,[P],Z_2)) \Big).
\end{multline}
Here $Z_1,Z'_1\in \Mor(F_1,F_1)$ are chains, $[P]\in \Mor(F_1,F_2)$ is a critical point of $F_1-F_2$,\; $Z_2\in \Mor(F_2,F_2)$ is another chain.
Here is a pictorial explanation of terms in $\int_{\dd\, MT}\PA$.
\begin{figure}[H]
$$
\hspace{-0.5cm}
\vcenter{\hbox{ \includegraphics[scale=0.8]{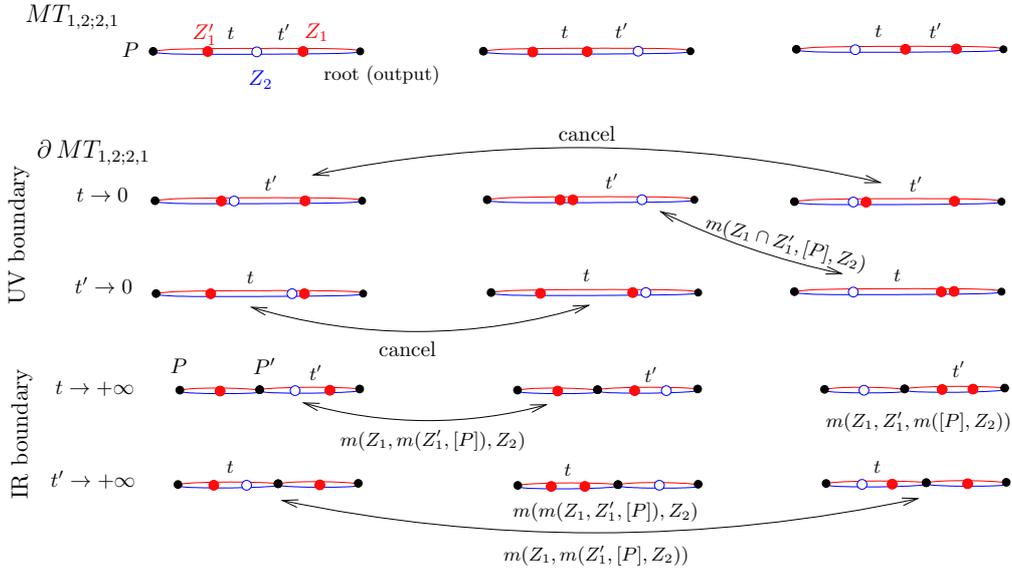} }}
$$
\caption{Contributions of boundary strata of the moduli space of metric trees to $\int_{\dd\, MT}I$ (and thus to an $A_\infty$ relation).}
\label{fig8}
\end{figure}
\end{example}

\section{Concluding remarks}
Inclusion of 
codimension-one cycles as in Section \ref{example: Novikov} may be interpreted as adding a flat connection (local system) to an object. 
One can actually consider replacing the object $F_a$ by $\nu_a$ copies of $F_a$ -- a ``stack'' of $\nu_a$ Lagrangian submanifolds $L_a$ in the language of Fukaya category of $T^*X$. This corresponds in the language of Section \ref{sec 3} to replacing the coefficient algebra of $N\times N$ matrices with algebra of block matrices with blocks of size $\nu_a\times \nu_b$.
In particular, this allows one to associate to an object $F_a$ a non-abelian flat connection on $X$ (or on $L_a$ in the language of $T^*X$) with coefficients in $GL(\nu_a)$.

Note that we have never used the concept of the  holomorphic disk neither in the first proof of $A_\infty$ relations (by effective action) nor in the second (via topological quantum mechanics).

We also remark that the construction of this paper can be generalized to the situation where differences of functions $F_a-F_b$ are only required to be Morse-Bott, not Morse -- thus, the critical locus $\mr{Crit}(F_a-F_b)$ is allowed to be a union of  submanifolds of different dimensions, rather than just a set of isolated points. Then one sets $\Mor(F_a,F_b)$ to be the singular chains on $\mr{Crit}(F_a-F_b)$, equipped with Morse-Bott differential which combines the boundary operator on chains with the data of gradient flows between critical manifolds. Note that in this approach one doesn't have to treat separately the case of morphisms from an object $F_a$ to itself -- it is an instance of the general construction (indeed, $\mr{Crit}(F_a-F_a)=X$, thus Morse-Bott chains in this case are just the singular chains on $X$).



%

\end{document}